# Spatially Resolved Thermometry of Resistive Memory Devices

Eilam Yalon[1], Sanchit Deshmukh[1,†], Miguel Muñoz Rojo[1,†], Feifei Lian[1], Christopher M. Neumann[1], Feng Xiong[1,2], and Eric Pop[1,3,4,*]

[1]*Department of Electrical Engineering, Stanford University, Stanford, CA 94305, USA.* [2]*Present address: Department of Electrical & Computer Engineering, University of Pittsburgh, Pittsburgh, PA 15261, USA.* [3]*Department of Materials Science & Engineering, Stanford University, Stanford, CA 94305, USA.* [4]*Precourt Institute for Energy, Stanford University, Stanford, CA 94305, USA.* [*]*E-mail: epop@stanford.edu*

**The operation of resistive and phase-change memory (RRAM and PCM) is controlled by highly localized self-heating effects, yet detailed studies of their temperature are rare due to challenges of nanoscale thermometry. Here we show that the combination of Raman thermometry and scanning thermal microscopy (SThM) can enable such measurements with high spatial resolution. We report temperature-dependent Raman spectra of $HfO_2$, $TiO_2$ and $Ge_2Sb_2Te_5$ (GST) films, and demonstrate direct measurements of temperature profiles in lateral PCM devices. Our measurements reveal that electrical and thermal interfaces dominate the operation of such devices, uncovering a thermal boundary resistance of 30 $m^2K^{-1}GW^{-1}$ at GST-$SiO_2$ interfaces and an effective thermopower 350 µV/K at GST-Pt interfaces. We also discuss possible pathways to apply Raman thermometry and SThM techniques to nanoscale and vertical resistive memory devices.**

[†] Authors contributed equally

## I. INTRODUCTION

Information storage and memory devices based on the change of resistance (i.e. resistive memories or memristors) hold several key advantages over contemporary charge-based memory devices. Such memory devices are two-terminal resistors that retain their resistance state as a function of the applied voltage or current. Several technologies can be included under the general term of "resistive memory"[1,2], such as phase change memory (PCM)[3,4], resistive random access memory (RRAM)[5,6], and conductive bridge RAM (CB-RAM)[7]. Memristive devices are important not only for memory and storage applications; they are also being extensively studied as computing elements for neuromorphic architectures[2,8].

The resistive switching in RRAM devices is based on the formation and rupture of conductive filaments in thin metal oxides, like $HfO_2$. In PCM, a nanoscale volume of chalcogenide material (like $Ge_2Sb_2Te_5$) can be SET to a crystalline (low resistive) state and RESET to an amorphous (high resistive) state using electrical pulses. Self-heating and the local temperature play a major role in the principle of operation of both PCM[4] and RRAM[9,10]. Many of the advantages (e.g. energy efficiency improved with scaling)[11,12] and shortcomings (e.g. reliability)[13] of these technologies stem from their inherent dependence on self-heating. Therefore, understanding the energy and heat dissipation mechanism is vital for the evaluation, design and optimization of all such future technologies. However, experimental techniques to measure nanoscale device temperature are challenging and scarce[10,14,15]. In particular, spatially resolved measurements that reveal the energy dissipation mechanism are required for better understanding of the device physics[16].

Here we examine for the first time the combination of Raman thermometry and scanning thermal microscopy (SThM) to measure the spatially resolved temperature rise in resistive memory devices. We present temperature-dependent Raman spectroscopy of thin films for two of the most commonly used RRAM oxides: $HfO_2$ and $TiO_2$, and of the PCM material $Ge_2Sb_2Te_5$ (GST). We further show an experimental measurement of the temperature profile in a Joule-heated PCM device, providing important insights into its operation. Finally, we discuss how Raman and SThM can be used to measure the local temperature rise in vertical and other nanoscale RRAM and PCM device geometries.

## II. Raman Spectroscopy of RRAM Oxide Films

Raman spectroscopy measures the shift in inelastically scattered light, directly corresponding to phonon energy ($\hbar\omega$) and temperature ($T$). Stokes (anti-Stokes) lines are due to photons scattered at lower (higher) energy than the incident laser, due to phonon emission and absorption, respectively. (More Raman spectroscopy details are given in Supplemental Information Section 1.) Fundamentally, most RRAM oxides have poor Raman signal due to weak absorption (ultra-thin films with large band gap) and low degree of crystallinity. However polycrystalline oxide film regions or the programmed RRAM filament could be expected to have different Raman signals, and are yet to be studied. For example, Raman spectroscopy has been previously used to study stoichiometry, defects, and particularly oxygen vacancies in crystalline[17] and nano-crystalline[18] oxides. The Raman spectra of oxide powders, single crystals, and thick films have previously been reported[19-21], but here we present the first temperature-dependent Raman spectra of nanoscale thin films which are relevant for RRAM devices.

Figure 1 shows the temperature-dependent Raman spectra of two of the most common RRAM oxides: (a) $HfO_2$ and (b) $TiO_2$. The 50 nm thick films were sputtered onto Pt/sapphire and $SiO_2$/Si substrates, respectively, and did not show any measurable Raman features in their (as-deposited) amorphous state. After annealing (see Methods) the films crystallized, exhibiting Raman signals of the monoclinic ($HfO_2$) and anatase ($TiO_2$) phases. The insets show the temperature dependence of a selected mode. The $B_{g1}$ monoclinic $HfO_2$ mode[22] (~134 cm$^{-1}$ at 25 °C) shows typical frequency downshift with temperature at rate of ~0.011 cm$^{-1}$/C. The $E_g$ anatase $TiO_2$ mode (~141 cm$^{-1}$ at 25 °C) however shows anomalous frequency increase with temperature. This trend was previously reported and explained via strong contribution of the quartic anharmonicity[21]. For the practical purpose of device thermometry it is sufficient to have a well-defined temperature response of the Raman mode, either positive or negative.

We note that temperature-dependent Raman data are presented in this section for nanoscale oxide films for the first time. However, obtaining sufficient signal in nanoscale RRAM *devices* is challenging, but could be addressed using signal enhancement techniques such as surface- or tip-enhanced Raman spectroscopy (SERS[23] or TERS[24,25]). In addition, an optically transparent electrode such as graphene[26], indium tin oxide (ITO), or a very thin (e.g. sub-10 nm) metal can be used to measure vertical device structures. In the next section we focus on Raman thermometry of PCM which exhibits a sufficiently strong signal to demonstrate (spatial) device thermometry.

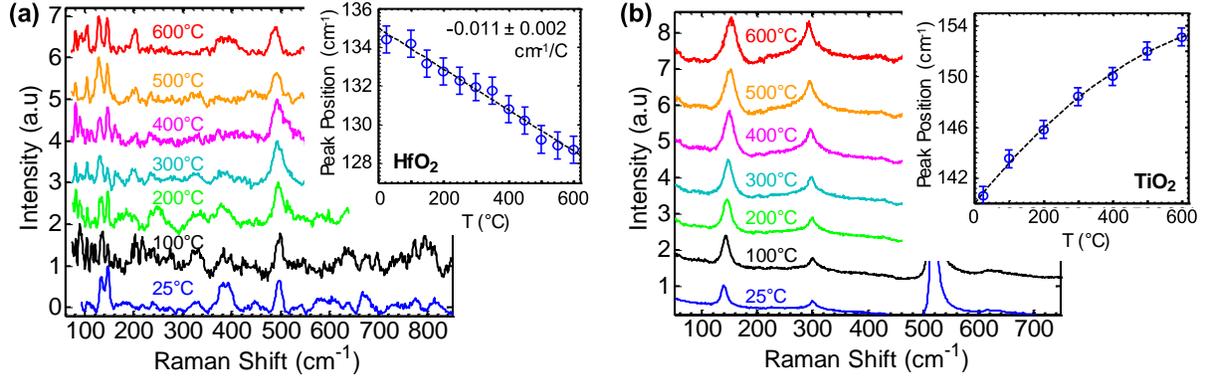

**Fig. 1**. Temperature dependent Raman spectra of RRAM oxide nanoscale thin films. Raman spectra of crystallized 50 nm thin films of **(a)** $HfO_2$ and **(b)** $TiO_2$, measured at temperatures ranging from 25 °C to 600 °C. Spectra are vertically offset for clarity. Insets show peak position shift with temperature of a selected mode. The $HfO_2$ was sputtered onto Pt (50 nm) on sapphire substrate and then annealed at 600 °C for 2 hours. All measured peaks above 110 cm$^{-1}$ are assigned to monoclinic phase of $HfO_2$ (as confirmed by XRD)[22]. The $TiO_2$ was sputtered onto $SiO_2$ (90 nm) on Si substrate and annealed at 400 °C for 1 hour. The peak at ~141 cm$^{-1}$ (25 °C) is assigned to anatase $TiO_2$ and the other peaks are from the Si substrate.

## III. RAMAN Spectroscopy of PCM

### A. Raman Spectroscopy of GST Films

Raman spectroscopy has previously been used to characterize nanoscale PCM films[27-29]. Unlike oxides, GST absorbs photons in the range of visible excitation lasers (~500-650 nm) due to its smaller band gap (~0.6 eV). In addition, much of the GST volume in a typical PCM cell is crystalline, thus Raman thermometry can be readily applied to PCM devices. Moreover, Raman spectroscopy can be used to identify phase and presence of defects in GST as outlined below.

Figure 2 shows, for the first time, the temperature dependent Raman signal of 20 nm thin sputtered GST films (see Methods). Figure 2(a) shows the first heating cycle on a hot stage from 25 °C to 400 °C. The as-deposited film is amorphous (a-GST) and starts to crystallize (c-GST) at

~140 °C, first to the face centered cubic (fcc) phase and then at ~240 °C to the hexagonal closest pack (hcp) phase. For assignment of the various Raman peaks to GST modes please see ref. [27]. Upon cooling to room temperature and heating back up to 450 °C the film remains in its stable hcp phase, as shown in Figure 2(b). Figure 2(c) shows the measured (symbols) and fitted peak of a selected mode at stage temperatures of 25 °C (blue, ~173 cm$^{-1}$) and 400 °C (red, ~168 cm$^{-1}$) and Fig. 2(d) shows the peak position downshifting vs. stage temperature from 25 °C to 450 °C.

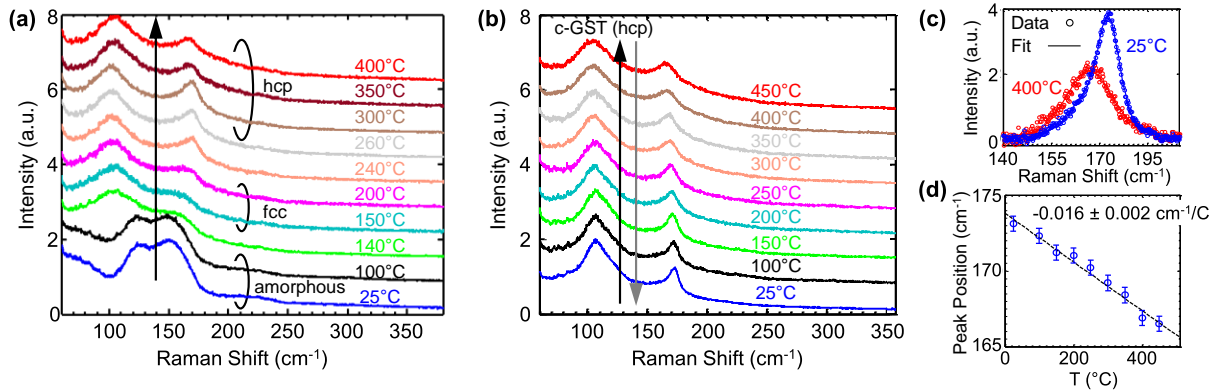

**Fig. 2**. Temperature dependent Raman spectra of 20 nm thin GST film capped in situ with 20 nm SiO$_2$ (see Methods). **(a)** Raman spectra during the first heating cycle on a hot stage from 25 °C to 400 °C. As-deposited, the GST is amorphous, but after heating the film starts to crystallize first to the fcc phase and then to the hcp phase. **(b)** Temperature dependent Raman spectra of the stable hcp phase GST at temperatures ranging from 25 °C to 450 °C. Similar spectra were obtained for cooling (not shown). **(c)** Example of measured (symbols) and fitted (lines) peak for a selected mode at 25 °C and 400 °C in the hcp phase. **(d)** Peak position shift vs. temperature of the selected mode shown in (c).

### B. Raman Spectroscopy of GST Devices

Thanks to its material and phase selectivity, Raman spectroscopy can be used to map GST films and PCM devices for phase analysis. However, here we uncover that patterned and processed GST *devices* exhibit Raman spectra that are different from blanket deposited GST *films*. Figure 3(a) schematically shows the Raman measurement of a lateral GST device with Pt contacts, and device fabrication details are provided in the Methods section.

Figure 3(b) shows the Stokes and anti-Stokes Raman signal of the GST device on a hot stage from 25 °C to 125°C. The stage temperature was kept below the highest temperature during processing (PMMA bake at 180 °C) in order to avoid changes to the GST channel and/or contacts. It is evident that the Raman signal of the GST device is dominated by two intense peaks at ~120 and ~140 cm$^{-1}$, which are not present in the GST film spectra (Fig. 2). These peaks were present in the GST device immediately after lift-off, before spin-coating the PMMA capping. The intensity of these peaks is significantly larger than the GST film peaks shown in Fig. 2.

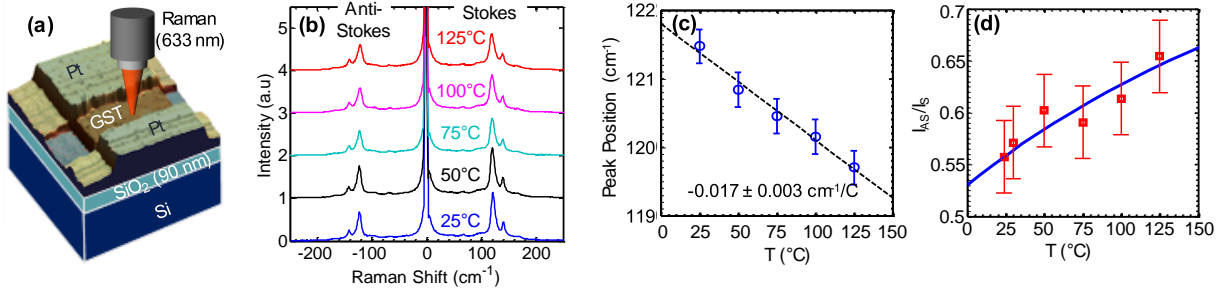

**Fig. 3**. Temperature dependent Raman spectra of lateral GST device. **(a)** Device and measurement setup. The GST channel ($W = 10$ µm, $L = 5$ µm) is patterned on top of Pt electrodes and capped with PMMA (see Methods). **(b)** Stokes and anti-Stokes Raman spectra of patterned GST device on a hot stage at temperatures from 25 °C to 125 °C. These are dominated by Te modes[30] with much higher intensity than GST modes. **(c)** Peak shift with temperature of a selected mode (at ~120 cm$^{-1}$). **(d)** Anti-Stokes to Stokes intensity ratio vs. temperature of a selected mode.

These intense peaks (~120 and ~140 cm$^{-1}$) are known from previous studies as Te peaks[30]. We have also measured similar peaks in uncapped GST films after exposure to high laser power and formation of dark spots in the film (Supplemental Information Section 2). These peaks were also measured in other sets of lateral GST devices which were capped with different oxides[31]. Such oxide-capped devices switched for many cycles (~10$^5$) with good on/off ratio (~10$^2$) and low switching energy (< 20 pJ). These Te peaks are likely associated with surface oxidation of the GST during processing, resulting in formation of GeO$_x$ and SbO$_x$, and precipitation of Te. The former is also evident from XPS of the dark spots at the GST surface (Supplemental Information Section 2), while the latter is evidenced in our measured Raman spectra.

The intense Te peaks have strong temperature dependence, as shown in Fig. 3(c) which allow us to use them efficiently as a thermometer. Raman spectroscopy can then be used to extract temperature in two different ways; one via the calibration of peak shift vs. temperature on the hot stage[32,33] and the other directly from the anti-Stokes to Stokes (AS/S) intensity ratio[34]:

$$T = \frac{\hbar \omega_{ph}}{k_B} \left[ 3\ln\left(\frac{\omega_L + \omega_{ph}}{\omega_L - \omega_{ph}}\right) - \ln\left(I_{AS}/I_S\right) \right]^{-1} \quad (1)$$

where $T$ is the phonon temperature (in K), $\omega_{ph}$ and $\omega_L$ are the phonon shift and laser frequency, respectively, $I_{AS}/I_S$ the anti-Stokes to Stokes intensity ratio, and $k_B$ is the Boltzmann constant. The advantage of the AS/S method is that in principle it does not require a calibration procedure and that the temperature can be obtained with a single measurement. This is attractive for spatial temperature mapping of devices since the temperature map can be obtained in a single map scan, whereas for the peak shift method a calibration measurement is needed and at least two map scans must be obtained[33] (e.g. one with and the other without bias).

The symbols in Fig. 3(d) show the AS/S intensity ratio of the spectra from Fig. 3(b) after baseline subtraction and Lorentzian peak fitting. The blue line represents eq. **1)**). The main drawback of the AS/S method is the uncertainty in the temperature evaluation, because the extraction of intensity from peak fitting is less accurate than the extraction of peak position. Moreover, the *absolute* temperature $T$ ($\geq$ 300 K) is obtained from the intensity ratio with relative error ~15% in our case. Since we are interested in measuring the *change* in temperature ($\Delta T$ ~ 100 K above room temperature) the relative error in $\Delta T$ is three times larger, or nearly ~ 50%. At the same time, the relative error in $\Delta T$ for the peak shift method is less than ~15% [e.g. see Fig. 3(c)]. Therefore, in the remainder of this study we use the peak shift method to map the temperature of the device, following the procedure outlined in Ref. 33 for aligning the two maps. We note that temperature sensitivity of the AS/S method should improve for phonon modes at higher frequencies (Supplemental Information Section 3).

## IV. PCM DEVICE THERMOMETRY

### A. Measurement Technique

Figure 4(a) displays the schematic of the Raman thermometry measurement applied to a lateral GST device. We spatially mapped the device with 0.25 µm step size, measuring the Raman spectra at each point with and without electrical bias. The temperature is extracted by converting the peak shift (with electrical bias and self-heating) to temperature via the calibration shown in Fig. 3(c). Figure 4(b) shows the measured (symbols) and fitted (lines) Raman spectra at the center of the GST channel with (red) and without (blue) electrical bias. The Te peaks at ~120 and 140 cm$^{-1}$ are clearly visible, as well as their shift with self-heating. Raman features at ~ 105 and 170 cm$^{-1}$ may correspond to GST hcp phase, but their signal is significantly smaller.

We also carried out scanning thermal microscopy (SThM) measurements on the same devices. SThM is an atomic force microscopy (AFM) based technique that uses a thermo-resistive probe to acquire nanoscale topographic and thermal images simultaneously[35,36]. Unlike Raman, SThM measures only the surface device temperature and requires additional calibration (see Supplemental Information Section 4), however it provides nearly AFM-like spatial resolution (<100 nm) and is therefore used here to provide complementary insight into PCM device thermometry.

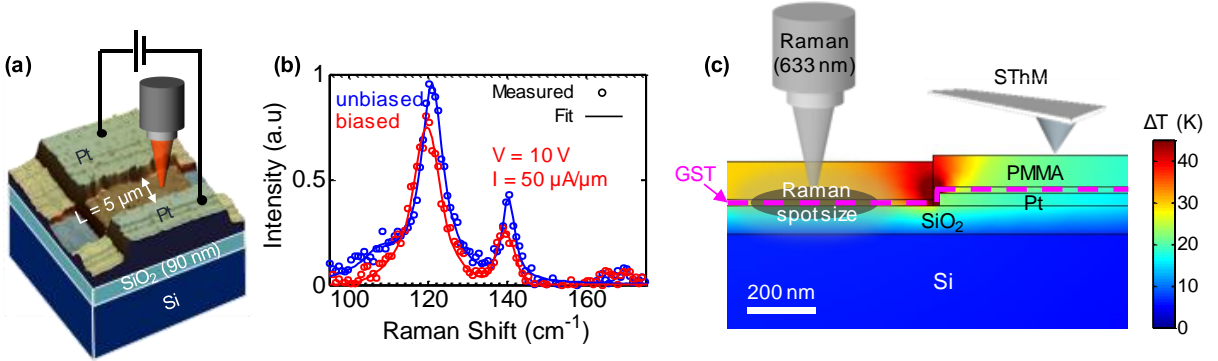

**Fig. 4.** PCM device thermometry on a lateral device. **(a)** Measurement setup: Raman (and SThM) acquired during device operation with self-heating. **(b)** Measured (symbols) and fitted (lines) Raman spectra of the GST at the center of the channel with electrical bias (red: $V = 10$ V, $I = 0.5$ mA) and without bias (blue). **(c)** Simulated cross-section temperature profile of the device highlighting the temperature measured by Raman (directly GST film with Gaussian laser spot size) and SThM (top PMMA surface).

Figure 4(c) shows the simulated cross-sectional temperature profile in our lateral GST device and contact. Three-dimensional (3D) finite element simulations were carried out using COMSOL Multiphysics ®. To better understand the measured temperature, we illustrate the Raman and SThM instruments with the cross-sectional simulated temperature. The Raman measures the temperature of the GST channel but its signal is averaged across the Gaussian laser beam spot (here ~ 400 nm). Inherently the SThM has better spatial resolution with a thermal exchange radius of < 100 nm[37], yet it measures the temperature at the top surface of the PMMA capping layer which spreads the heat. Simulations therefore predict slightly different temperature profiles for the Raman and SThM measurements. We note that the SThM is operated here in DC mode and is not calibrated to output temperature directly, rather it outputs a relative signal. We then use the temperature measured by Raman (with same power input conditions) to calibrate the SThM signal to the temperature of the PCM device center, where the Raman and SThM temperatures are expected to be similar [see simulated cross-sectional temperature rise in Fig. 4(c)].

The main advantage of the Raman measurement is its material selectivity which allows a differential measurement of materials in the laser path, thus enabling even atomic scale resolution in the cross-plane direction[38]. However, not all materials have a usable Raman signal and the spatial resolution is diffraction-limited (unless a signal enhancement technique such as TERS[24] is utilized). SThM on the other hand has nearly AFM-like spatial resolution, but measures the temperature at the top surface rather than the direct temperature of the material of interest. The temperature sensitivity of the SThM can be better than that of the Raman, but in order to convert the SThM signal (voltage) to temperature, the Raman was used for calibration, as outlined above. Both techniques are limited in measuring *vertical* RRAM devices; the top electrode might block the Raman signal and laterally spread the temperature profile measured by SThM. Nonetheless, the combination of both techniques could become a very powerful tool to study the power dissipation in cases where temperature plays a major role in device operation[2,3,5,10]. A practical solution is to measure lateral devices or vertical devices having transparently thin top electrode.

### B. Power Dissipation in Lateral PCM Device

Figure 5 compares the temperature profile along the device channel (including the contacts) measured by (a) SThM and (b) Raman to the predicted temperature rise with very good agreement. The full device temperature maps are shown in Supplemental Information Sections 5 and 6. Figure 5(c) shows the simulated temperature rise of the GST channel (green) as well as the profile that would be measured by SThM (red) and Raman (blue). Figure 5 highlights how the power dissipation in our devices is revealed via both Raman thermometry and SThM, with varying degrees of spatial accuracy. By inspecting the measured temperature profiles in Fig. 5 and the fitting parameters used in the simulation (summarized in Table I of Supplemental Information Section 9), several conclusion can be drawn as follows.

First, it is evident that significant heat is generated at the contacts[39]. It appears that the large contact resistivity of the GST film on the Pt contact (GST is deposited on top of the Pt electrodes) lead to highly localized power density, mainly at the edge of the electrode (also evident in the SThM maps shown in Supplemental Information Section 5). We also carried out transfer length method (TLM) measurements to directly extract the contact and sheet resistance (Supplemental Information Section 7) which served as inputs to the simulations. However, the temperature peak at the Pt-GST edge suggests that much of the contact voltage may be dropped at the imperfect GST step coverage of the Pt electrodes. To account for this, we set the GST resistivity at the Pt sidewall to be 20x larger than the bulk GST resistivity, leading to correct fitting in our simulations. The temperature rise of the 5 µm long GST channel suggests that much of the power dissipates there, yet the temperature peaks at the contacts due to the high power *density* there.

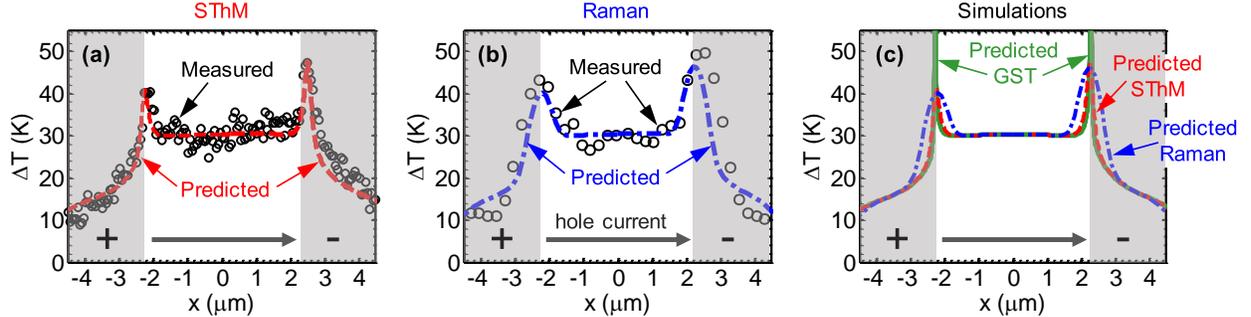

**Fig. 5.** Temperature rise along GST channel in fcc phase (black symbols) measured by **(a)** SThM and **(b)** Raman thermometry. The gray zones mark the contact regions. The temperature rise predicted by finite element simulations for SThM (red dashed line) and Raman (blue dash-dot line) are also shown in (a) and (b) respectively. **(c)** Simulated temperature rise of the GST channel (green solid line), the SThM (top of PMMA surface averaged across thermal exchange radius[37]) same as in (a) and Raman (GST with Gaussian laser spot size) same as in (b).

Second, given the power input and the measured temperature rise we can estimate the thermal boundary resistance (TBR) of the GST-SiO$_2$ and GST-Pt interfaces to be ~30 and ~ 20 m$^2$K/GW respectively. These values are in agreement with the TBR of the GST-SiO$_2$ interface previously measured[40] by the time domain thermo-reflectance (TDTR) technique, but they are extracted here within a functioning PCM device for the first time. The measured TBR of the GST-SiO$_2$ interface is equivalent to a Kapitza length ~50 nm of SiO$_2$, which accounts for more than 25% of the total thermal resistance in our device. Importantly, the relative contribution of the TBR is expected to increase and dominate as devices are scaled down in size[41].

Third, we also observe a clear asymmetry in heating at the two contacts, with the higher temperature at the edge of the grounded electrode. This contact heating asymmetry is due to a combination of thermoelectric and thermionic effects[42,43] as confirmed by reversing the current flow direction. We extract an effective thermopower $S$ ~ 350 µV/K, which includes both thermoelectric and thermionic effects at the GST-Pt contact. Similar values were reported for GST in the mixed amorphous and fcc phase with Pt[44] and with TiW[45] slightly above room temperature, after annealing up to ~ 150 °C. Following anneal at higher temperature (> 200 °C) and higher degree of crystallization the thermopower is expected to drop significantly (below 50 µV/K)[44,45]. The higher temperature at the grounded contact is also consistent with the location of damage after device breakdown, shown in Supplemental Information Section 8. The asymmetric heating highlights the importance of designing the voltage polarity of PCM programming pulses, to take advantage of thermionic and thermoelectric effects[46]. We note that our test devices are lateral, and larger than state-of-the-art PCM devices[47-50]; however the physical insights are valid and highly valuable: the power generation is dominated by electrical contacts and the heat dissipation is limited by thermal interfaces in nanoscale devices[39].

## V. CONCLUSION

In summary, in this study we have laid the fundamental basis for thermometry of RRAM and PCM data storage devices. We presented the first measurements of thermal Raman signatures in nanoscale films of $HfO_2$ and $TiO_2$, and we used spatial mapping of temperature (with both Raman and SThM) to provide physical insight into the operation of PCM devices. Our approach takes advantage of the benefits of each technique, e.g. selectivity (Raman) and high spatial resolution (SThM), and can be extended to a wide variety of devices. We uncover significant heating at the contacts, suggesting that power dissipation is often dominated by electrical contact resistance. Contact heating is asymmetric, depending on current flow direction, showing that thermoelectric effects must be taken into account when designing PCM programming pulses. We also extract the TBR of the GST-$SiO_2$ interface and find that it is equivalent to ~ 50 nm thick $SiO_2$, contributing much of the device thermal resistance. The role of both electrical contacts and thermal interfaces will only become more dominant as devices are scaled to sub-50 nm dimensions. Uncovering the spatial distribution of temperature rise in such self-heated memristive devices is essential for their understanding, and their future design and integration.

**Methods**

Raman measurements were carried out on a Horiba Labram Evolution HR using a 633 nm laser with an 1800 l/mm grating. The red laser is chosen here since it provides Raman signal comparable to other laser lines in our system (e.g. 532 nm) and allows the measurement of anti-Stokes signal by using a volume Bragg grating optical filter at 633 nm.

HfO$_2$ films were deposited by reactive sputtering from an Hf target in an Ar:O$_2$ (7:3) plasma at 4 mTorr, with a forward RF power of 150 W, at room temperature onto thin Pt films (50 nm) on a sapphire substrate. The Pt/sapphire substrate was used to enhance the Raman signal of the HfO$_2$ film. TiO$_2$ films were sputtered from a Ti target in an Ar:O$_2$ (14:1) plasma at 5 mTorr onto SiO$_2$ (90 nm) on Si substrates. To obtain Raman signal the films were crystallized by annealing: the TiO$_2$ for 1 hour at 400 °C and the HfO$_2$ for 2 hours at 600 °C, both in air.

Blanket GST films (20 nm) discussed in section III.A were sputtered onto SiO$_2$ (90 nm) on Si substrates, immediately followed (without breaking chamber vacuum) by sputtering 20 nm SiO$_2$ to prevent oxidation when later exposed to air. GST devices (section III.B) were prepared as follows. First, contacts and pads were defined by photolithography. Contact separation defined channel lengths $L$ = 2 to 20 µm. The 40 nm Pt contacts (with 2 nm Ti adhesion layer) were then deposited by e-beam evaporation followed by patterning of channels (of widths $W$ = 1 to 10 µm) with e-beam lithography, sputtering of 20 nm GST and lift-off. Devices were capped by spin coating ~150 nm poly(methyl methacrylate) (PMMA) to prevent oxidation, then baked on a hot plate at 180 °C for 10 minutes (in air) to crystallize the GST film.

The SThM measurements were carried out in passive mode and under DC bias. Measured PCM devices were Joule heated electrically by applying constant voltage to the contact pads for ~ 10 minutes during the SThM scan. The SThM thermal probe model used in this work is a DM-GLA-5 from Anasys®, which is made of a thin Pd layer on SiN.

**Acknowledgements**

We thank Ajay Sood, Scott Fong and Ilya Karpov for helpful discussions. We acknowledge the Stanford Nanofabrication Facility (SNF) and Stanford Nano Shared Facilities (SNSF) for enabling device fabrication and measurements. This work was supported in part by the NSF Center for Power Optimization of Electro-Thermal Systems (POETS), by the NSF grant DMREF 1534279, by DARPA Matrix program (HRL Laboratories subcontract), and by the Stanford Non-Volatile Memory Technology Research Initiative (NMTRI). E.Y. acknowledges partial support from Ilan Ramon Fulbright Fellowship and from Andrew and Erna Finci Viterbi Foundation.

# Spatially Resolved Thermometry of Resistive Memory Devices

Eilam Yalon[1], Sanchit Deshmukh[1,†], Miguel Muñoz Rojo[1,†], Feifei Lian[1], Christopher M. Neumann[1], Feng Xiong[1,2], and Eric Pop[1,3,4,*]

[1]*Department of Electrical Engineering, Stanford University, Stanford, CA 94305, USA.* [2]*Present address: Department of Electrical & Computer Engineering, University of Pittsburgh, Pittsburgh, PA 15261, USA.* [3]*Department of Materials Science & Engineering, Stanford University, Stanford, CA 94305, USA.* [4]*Precourt Institute for Energy, Stanford University, Stanford, CA 94305, USA.* [*]*E-mail:* [epop@stanford.edu](epop@stanford.edu)

[†]*Authors contributed equally*

## Supplementary Information

Table of Contents

## S1. Raman spectroscopy

Raman spectroscopy measures the shift in inelastically scattered light, directly corresponding to phonon energy ($\hbar\omega$) and temperature (T). Stokes (anti-Stokes) lines are due to photons scattered at lower (higher) energy than the incident laser, due to phonon emission and absorption, respectively. Thus, the Raman scattering process directly probes the phonon modes in the sample and provides information on the atomic structure, while also being sensitive to strain[1], doping[2], defects[3], and temperature[4,5]. The sensitivity to temperature arises from phonon mode softening (peak position downshift) as well as increased phonon scattering (peak broadening) with increasing temperature. In addition, the anti-Stokes to Stokes intensity ratio depends on the phonon population which is directly related to temperature[6]. The capability of measuring nanoscale features (such as individual carbon nanotubes)[7] combined with the sensitivity to temperature result in a potentially attractive technique for RRAM and PCM thermometry. Other details of the Raman measurements carried out in this work are presented in the Methods section of the paper.

## S2. Raman signal of patterned GST devices

We studied the origin of the Raman features found in our patterned GST devices which show a different spectra compared with blanket deposited GST (see Figs. 2 and 3 in the main text). The Raman spectra of our patterned GST devices is shown in Fig. S1a (red) and includes two prominent features: one at ~122 cm$^{-1}$ and the other at ~142 cm$^{-1}$ which are clearly associated with the presence of Te (Fig. S1e-f)[8]. We compare the patterned GST Raman spectra with that of other GST films that most likely oxidized during laser-induced heating (Fig. S1b) and during electrical heating (Fig. S1d) as well as uncapped $Sb_2Te_3$ (partially oxidized at the surface, Fig. S1c). These GST and $Sb_2Te_3$ films were also measured by XPS and show evidence of surface oxidation (Fig. S2).

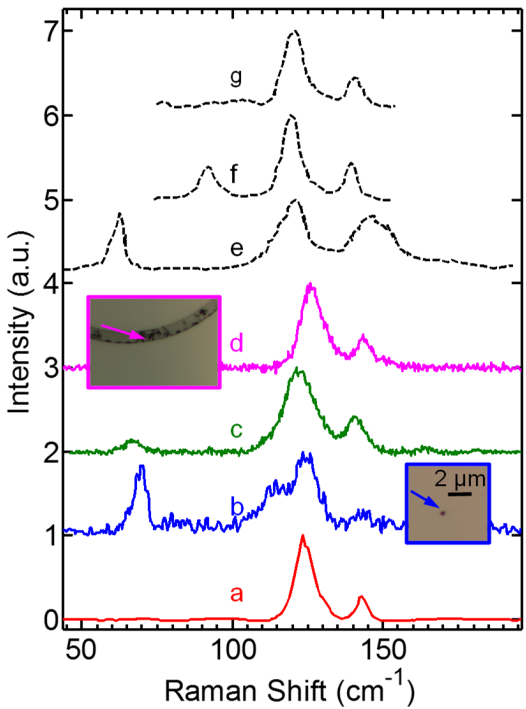

**Figure S1 | Raman signatures of oxidized GST, $Sb_2Te_3$ and Te.** Raman spectra of: (a) Patterned and finalized GST device described in the main text, after lift-off, (b) dark spot formed in uncapped GST film (inset, blue) following Raman measurement at high laser power (~10 mW). (c) $Sb_2Te_3$ uncapped film, (d) dark spot formed in uncapped GST circular device (inset, red) after showing resistance changes induced by electrical pulses, (e) amorphous Te[8], (f) trigonal z-axis of Te[8], and (g) trigonal x-axis of Te[8].

Figure S2 shows X-Ray photoelectron spectroscopy (XPS) data of uncapped GST device (20 nm) thick) before and after a short Ar sputtering etch. The pre-sputter results represent the GST surface. Both Ge and Sb show native oxide peaks, whereas Te does not. Post-sputtering (i.e. a few nm into the film) the $Sb_2O_5$ peak decays, suggesting the native oxide is only present at the top few nm of the film. We note that the sputtering can also induce some damage to the film, so these results are mostly qualitative. Overall, Fig. S1 and Fig. S2 suggest that surface oxidation of GST results in formation of native Ge and Sb oxides as well as Te precipitates. These Te precipitates dominate the Raman signal.

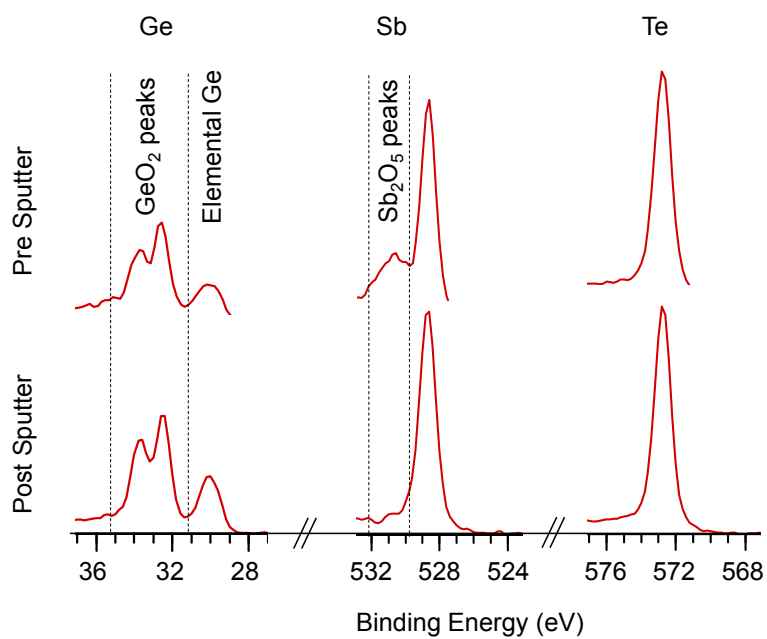

**Figure S2 | XPS of uncapped (oxidized) GST.** Pre-sputter (top panel) represents the GST surface, showing features of native Ge and Sb oxides as well as (un-oxidized) Te. The post-sputtering measurement (bottom panel) probes a few nm into the film, but might induce damage at the surface. The $Sb_2O_5$ feature decays post-sputtering, suggesting Sb is oxidized only at the very top surface.

## S3. Frequency dependence of the anti-Stokes to Stokes intensity ratio

The temperature in our experiment is measured by comparing the Raman *peak shifts* in an electrically biased sample to a calibration measurement on a hot stage. The anti-Stokes to Stokes (AS/S) Raman intensity ratio can also be used to probe temperature, without the need for a calibration measurement (on a hot stage) and/or a reference measurement (without electrical bias). In our experiment the uncertainty in temperature measured by $I_{AS}/I_S$ is significantly larger than the one measured by the peak shift method. One reason is that the uncertainty in measured (and fitted) intensity is larger compared with the measurement and fitting of (spectral) peak position. In addition, we illustrate in Fig. S3 that the temperature sensitivity in the AS/S method is poor for low wavenumbers (in our experiment the peaks are located at ~120 and 140 $cm^{-1}$). We note that for materials with Raman modes at higher frequency, the AS/S method could potentially be used with lower uncertainty. On the other hand, the AS signal intensity degrades with increasing mode frequency due to the decrease in phonon absorption probability.

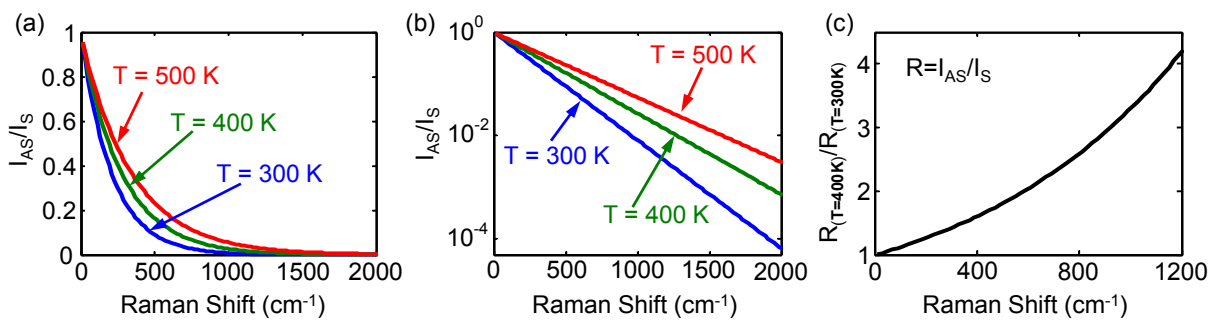

**Figure S3 | Anti-Stokes to Stokes ($I_{AS}/I_S$) signal sensitivity.** Calculated $I_{AS}/I_S$ vs. Raman shift (peak position) at T = 300 K (blue), 400 K (green), and 500 K (red) in (a) linear and (b) log y-axis. The temperature sensitivity improves with increasing wavenumber, but the $I_{AS}$ signal intensity decreases. (c) Sensitivity to temperature rise of ~ 100 K above room temperature; plot shows the ratio between R = $I_{AS}/I_S$ at 400 K and 300 K vs. Raman shift of the measured mode.

## S4.  Scanning thermal microscopy (SThM)

Scanning thermal microscopy (SThM) is an atomic force microscopy (AFM) based technique that uses a thermo-resistive probe to acquire nanoscale topographic and thermal images simultaneously[9,10]. When the probe is brought into contact with the sample surface, heat is exchanged between the surface and the probe. These temperature changes are correlated with the electrical resistance of the probe, which is measured using a Wheatstone bridge. The SThM has two main operating modes: one active, in which the probe is heated up and acts as both the heater and the thermometer, and the other passive, in which the probe is only used as the temperature sensor. The probe or sample heating can be carried out by DC or AC currents, which allow multiple ways of measuring and processing data. The SThM technique has been used to measure the thermal properties of films[11,12], nanowires[13,14], and devices[15]. In this study, the PCM device is Joule-heated in DC mode and the probe is used passively for sensing the local temperature rise.

## S5.  SThM images

Figure S4 displays SThM maps at varying input power and the corresponding AFM topography map. The heating maps show significant heating at the contact edges.

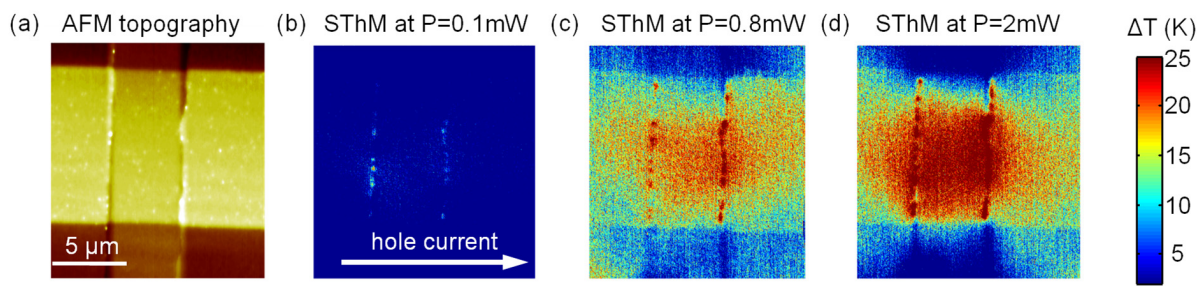

**Figure S4 | AFM and SThM.** (a) AFM topography map of the GST device described in the main text. Scanning thermal microscopy (SThM) maps of the same GST device with varying power inputs: (b) P = 0.1 mW, (c) P = 0.8 mW, and (d) P = 2 mW. The SThM heating map is calibrated via the Raman temperature measurement to obtain temperature values. Arrow in (b) indicates hole current flow direction for all 3 maps shown. The thermal maps show significant heating at the contact edges.

## S6. Temperature maps

We obtained temperature maps of the GST devices by Raman and SThM. Representative temperature profiles along the device (obtained by averaging temperature across device width) are shown in Fig. 4 of the paper. Fig. S5 compares measured and simulated temperature maps obtained by (b,c) SThM and (e,f) Raman. The (a) AFM topography map and (d) Raman intensity map are shown as reference to locate the channel and contact areas.

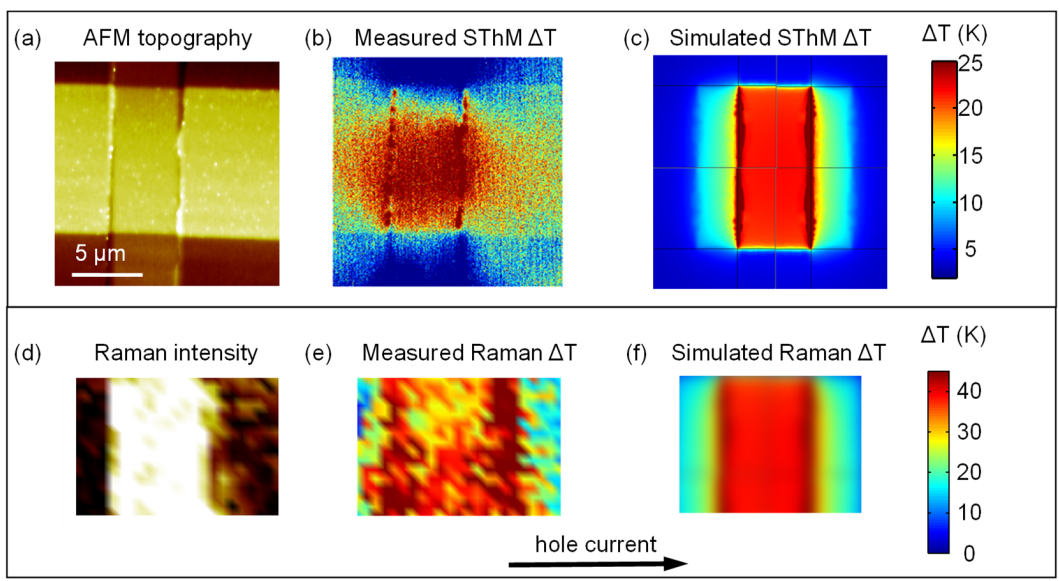

**Figure S5 | Temperature maps: SThM, Raman and simulations.** (a) AFM topography map of the GST device described in the main text. (b) Measured and (c) simulated SThM map of the same device at input power P = 2 mW. (d) Intensity map of the Raman mode at ~122 cm$^{-1}$. (e) Measured and (f) simulated Raman temperature map of the same device at input power P = 3.6 mW. Arrow indicates hole current flow.

## S7. Transfer length method (TLM)

Figure S6 displays TLM measurements of our GST films, showing the electrical resistance vs. channel length to obtain contact and sheet resistances. It is evident that in our lateral GST devices, for channels shorter than ~ 5 µm the contacts account for more than 50% of the total resistance, and therefore most of the power will be dissipated at the contacts. The extracted resistivity of the GST ρ ≈ 35 mΩ·cm agrees with previous studies[16] given the annealing temperature (180 °C for 10 minutes, leading to fcc phase) used here. The *contact resistivity* could not be reliably evaluated from the measured *contact resistance* due to the poor coverage of the GST film on the Pt contact sidewall and the resulting current crowding at the contact edge.

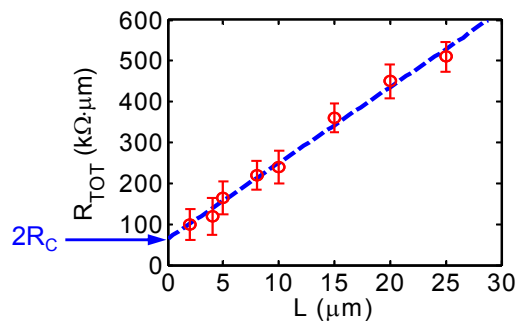

**Figure S6 | Transfer length method (TLM)**. Electrical resistance (normalized by width) of our GST test structures with varying channel length to obtain contact and sheet resistivity. The y-axis intercept represents the resistance of the contacts ($2R_C$), and the sheet resistance can be obtained from the slope.

## S8. GST device breakdown

The lateral GST devices undergo breakdown at high electrical bias, rather than reset to amorphous state. These devices degrade when the GST channel is self-heated, ostensibly because the PMMA capping layer evaporates, leaving the GST exposed to air at high temperature, at which it oxidizes. Optical images of the devices post-breakdown reveal damage to the GST film near the ground terminal, as shown in Fig. S7a. The location of the breakdown spot agrees well with the measured peak temperature (Fig. S7b).

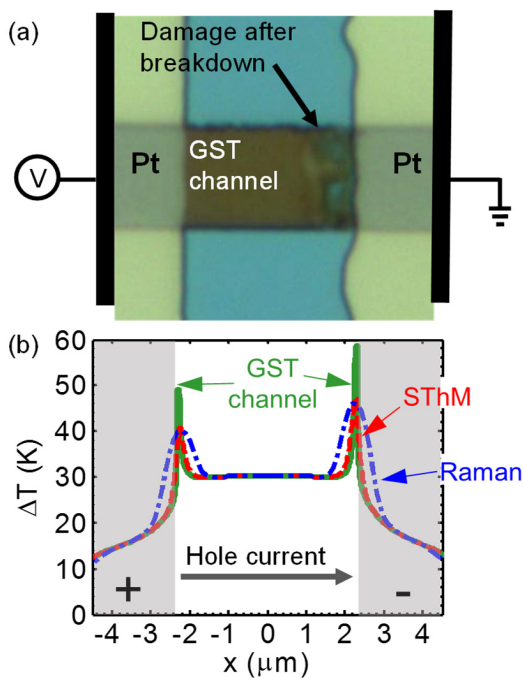

**Figure S7 | Post-breakdown damage**. (a) Optical image of GST channel after breakdown showing damage near the grounded electrode. (b) Simulated temperature profile (calibrated by measurements) of the GST channel (see Fig. 5 in the main text for details) showing the peak temperature located near the grounded terminal.

## S9. Parameters used in COMSOL simulation

Table I – parameters used in finite element modeling simulations (near room temperature)

| Parameter | Description | Value | Reference\comments |
|---|---|---|---|
| $\rho_{GST}$ | GST Electrical specific resistivity | 35 mΩ·cm | TLM measurement (fcc) |
| $\rho_{C,GST}$ | Electrical contact resistivity between GST-Pt | $4 \cdot 10^{-4}$ Ω·cm$^2$ | TLM measurement (fcc) |
| $\rho_{Pt}$ | Pt Electrical specific resistivity | $1.8 \cdot 10^{-7}$ Ω·cm | 4-probe measurement of similar films |
| $k_{GST}$ | GST thermal conductivity | 1 Wm$^{-1}$K$^{-1}$ | We set $k_{GST}$ = 1 Wm$^{-1}$K$^{-1}$, (Ref. 17) but the simulated temperature rise in the lateral device is relatively insensitive (less than 10% change) to thermal conductivity changes in the range $k_{GST}$ ~ 0.5 to 1.5 Wm$^{-1}$K$^{-1}$ |
| $k_{Pt}$ | Pt thermal conductivity | 50 Wm$^{-1}$K$^{-1}$ | Estimated from measured $\rho_{Pt}$ via Wiedemann Franz law |
| $k_{PMMA}$ | PMMA thermal conductivity | 0.19 Wm$^{-1}$K$^{-1}$ | COMSOL materials library. Steady state temperature profile is insensitive to this parameter |
| $k_{SiO2}$ | SiO$_2$ thermal conductivity | 1.4 Wm$^{-1}$K$^{-1}$ | Ref. 18 |
| $k_{Si}$ | Si (p$^{++}$) thermal conductivity | 95 Wm$^{-1}$K$^{-1}$ | Ref. 5 |
| TBR GST-SiO$_2$ | Thermal boundary resistance (TBR) of GST-SiO$_2$ interface | 30 m$^2$K/GW | Determined by fitting to experimental data ±5 m$^2$K/GW |
| TBR GST-Pt | TBR of GST-Pt interface | 20 m$^2$K/GW | Determined by fitting to experimental data ±5 m$^2$K/GW |
| TBR Si-SiO$_2$ | TBR of Si-SiO$_2$ interface | 3 m$^2$K/GW | Ref. 18 |
| TBR GST-PMMA | TBR of GST-PMMA interface | 20 m$^2$K/GW | Set to similar value as GST-Pt, but steady state temperature profile is insensitive to this parameter |

**Supplemental References**